\documentclass[aps,a4paper,preprint]{revtex4}
\usepackage{epsfig}
\usepackage{graphicx}
\usepackage{color}
\usepackage[english]{babel}
\usepackage{fancybox}
\usepackage{amsmath}
\usepackage{amsthm}
\usepackage{eucal}
\usepackage{amssymb}
\usepackage{mathrsfs}
\usepackage{mathtools}
\usepackage{natbib}
\def \beq {\begin{equation}}
\def \eeq {\end{equation}}
\def \beqn {\begin{eqnarray}}
\def \eeqn {\end{eqnarray}}
\usepackage{amsmath,amssymb,braket}

\begin{document}

\begin{flushright}

\end{flushright}

\title{On magnetic and vortical susceptibilities of the Cooper  condensate}

\author{  A. Gorsky $^{1,2}$ and F. Popov$^{2,3,4}$}

\address{$^{1}$Institute of Information Transmission Problems of the Russian Academy of Sciences,
Moscow, Russia, \\ 
$^{2}$ Moscow Institute of Physics and Technology, Dolgoprudny 141700, Russia\\
$^{3}$ Institute of Theoretical and Experimental Physics, Moscow, Russia\\
$^{4}$Department of Physics, Princeton University, Princeton, NJ 08544, USA.}

\begin{abstract}

We discuss the susceptibility of the Cooper condensate in the s-wave $2+1$ superconductor in the external
magnetic field and in the rotating frame. The 
extended holographic model involving the charged rank-two field is considered and it is
argued that the susceptibility does not vanish.  We interpret non-vanishing susceptibilities
as the admixture of the p-wave triplet  component in the Cooper condensate in the external
field.

\end{abstract}

\maketitle

\section{Introduction}

The ground state of conventional superconductors (SC)   involves the Cooper pairs 
in the different spin 
and orbital states forming the charged condensate. We shall be interested in the specific respond of the
ground state of $2+1$ dimensional s-wave superconductor to the external magnetic field and
rotation namely  if the p-wave component of the condensate proportional
to the external field is generated which is absent otherwise. The known examples 
of such phenomena one could
have in mind are the generation of the triplet component via the Rashba term \cite{rashba}
or via spin-orbit interaction \cite{spinorbit} in the singlet s-wave SC. 

The interesting possibility of coexistence of the triplet and singlet 
order parameters in the SC occurs in the specific materials
admitting the coexistence of the SC and antiferromagnetic (AF) orders.
In this case at least in the model description the following relation
takes place \cite{scaf}
\beq
|\Delta_t|\propto |\Delta_s| |M|
\eeq
where $\Delta_t, \Delta_s$ are the SC order parameters for triplet and singlet 
states while M is the AF magnetization order parameter. A bit loosely it can be claimed
that the AF component induces the triplet Cooper pairing. In what follows
our consideration has some similarities with this situation.

In the ground state of the conventional SC the magnetic field is screened by the  supercurrent and is expelled 
from the bulk   penetrating through the Abrikosov strings only.  However 
one can imagine that the triplet spin  component of the condensate can be generated by the external 
field in the bulk of the singlet SC. We 
shall look at the magnetic susceptibility of the Cooper condensate in the weak  magnetic field defined as 
\beq
<0|\psi\sigma_{\mu\nu}\psi|0> = g\chi_s <0|\psi\psi|0>F_{\mu\nu}
\label{sup}
\eeq
where g is dimensionful gauge coupling in $2+1$.
We will be interested if $\chi_s\neq 0$ and discuss this issue from the holographic viewpoint.

There
is the well-known analogy between the external magnetic field and the rotation
which is encoded in the specific form of external metric, see for example, the recent discussion in \cite{svistunov}.
Therefore it is natural to consider the responses of the ground state 
of superconductor at the external magnetic and gravimagnetic field  in parallel.
To this aim we shall also introduce and discuss the vortical susceptibility 
of the Cooper condensate in the rotating frame.

The partial motivation for this study is as follows.
Consider the hadronic phase in QCD where the chiral symmetry is broken by the
condensate $<\bar{\Psi}\Psi>$. The linear response of
chiral condensate to the external magnetic field is parametrized as follows
\beq
<0|\bar{\Psi}\sigma_{\mu\nu}\Psi|0> = \chi_F <\bar{\Psi}\Psi>F_{\mu\nu},
\label{sus}
\eeq
where $\chi_F$ is the magnetic susceptibility of the condensate introduced in \cite{ioffe}.
The value of $\chi_F$ can be derived by the different means. In particular  its value can be obtained from the anomalous 
CS terms in the conventional holographic model \cite{gk,gkkv} and in the extended model with additional 
rank-two fields \cite{harvey}. 

Recently the vortical susceptibility for the quark condensate was
introduced and evaluated in the dense QCD via specific anomaly in the dense matter \cite{afgk}
\beq
<0|\bar{\Psi}\sigma_{\mu\nu}\Psi|0> = \chi_G <\bar{\Psi}\Psi>G_{\mu\nu},
\label{sus}
\eeq
 where $G_{\mu\nu}$ is the curvature of the graviphoton field. The vortical
susceptibility is the response of the chiral condensate to the external 
gravitational field corresponding to the rotation frame.

Of course there  are some differences between QCD chiral condensate  and superconducting condensates.
The chiral condensate  is neutral while the Cooper condensate is charged. In QCD
the analogue of the Cooper condensate occurs only at high density in the color-flavor
locking superconducting phase while the chiral condensate corresponds to the neutral 
exciton condensate in the condense matter context. Let us emphasize that in 
the superconductor case  contrary 
to QCD we deal with the non-relativistic system.

In this Letter we consider the  susceptibility  of the SC $\chi_s$ 
combining the conventional and
holographic means. To this aim we consider the 2+1 SC described by the $AdS_4$ 
bulk geometry in the extended holographic model which involves complex scalar, U(1) gauge field
and rank-two field. We argue that both magnetic and vortical susceptibilities do not vanish.

The paper is organized as follows. In Section 2 we remind the simplest holographic models
of s-wave superconductor. In Section 3 we introduce the polarization of the Cooper condensate
in the magnetic field. Section 4 involves the arguments showing that the magnetic susceptibility does not vanish
for $2+1$ and $3+1$ cases.
In Section 5 we make a few comments concerning the vortical susceptibility of the condensate.
The results and open questions are summarized in the Conclusion.

\section{The Holographic model}	

	In this section we discuss the dual model for 2+1 superconductor. In the case of s-wave  superconductivity the relevant dual model 
	reads as (see \cite{rev1,rev2,zaanen} for reviews)
	\beqn	
	S = \int d^4 x \sqrt{-g}\left[R + \frac{\sigma}{L^2} -  F_{\mu\nu}^2 - \left|\partial_\mu \Psi - i g A_\mu \Psi\right|^2 - m^2 \left|\Psi\right|^2\right],\\
	ds^2 = - f dt^2 + \frac{dr^2}{f} + r^2\left(dx^2 + dy^2\right), f= \frac{r^2}{L^2}\left(1 - \frac{r_0^2}{r^2}\right).
	\eeqn 
We work in the rigid background space-time - AdS with black hole, and do not consider the feedback on the gravity by scalar $\Psi$ and electromagnetic field $A_\mu$. The charged scalar $\Psi$ is dual to the condensate $<\psi \psi>$ in an usual s-wave superconductor and $A_\mu$ is dual to the electric current. In the vicinity of the boundary we have the following asymptotic behavior of the fields.
	\beqn
	\Psi = \frac{\Psi_1}{r} +\frac{{\Psi_2}}{r^2}+ \dots\\
	\Phi = A_0 = \mu - \frac{\rho}{r},\qquad
	A_x = B y  + \frac{J_x}{r},\quad J_x = 0.
	\eeqn
	where the zero-component of electromagnetic field provides the chemical potential $\mu$ and charge density of dual theory on the boundary. The dimension of $\left[<\psi\psi>\right] = 3, \left[\Psi\right] = 1$ and  $\Psi_2$ corresponds to the value of condensate $<\psi\psi>$. 
	
	We study the behavior of $J_{\mu\nu} = <\psi \sigma_{\mu\nu}\psi>$ in the presence of external magnetic field $A_x = B y$ and  introduce an antisymmetric  field $B_{\mu\nu}$ that will be a source for charged tensor current $J_{\mu\nu}$.
	The Lagrangian for  antisymmetric field $B_{\mu\nu}$ has the following form
	\beq\label{add}
	\Delta L = \left|dB - i g A\wedge B\right|^2  - m^2 \left|B_{\mu\nu}\right|^2  + \lambda \Psi^\dagger B_{\mu\nu} F^{\mu\nu} + \lambda \Psi B^\dagger_{\mu\nu}F^{\mu\nu} .
	\eeq

It is useful to compare the Lagrangian (\ref{add}) with the Lagrangian for the antisymmetric tensor field
considered in the extended holographic model for QCD \cite{karch,harvey}. Remind that the minimal holographic QCD
model defined in 5d AdS-like space involves the gauge fields $A_L,A_R$ in  $ U(N_F)\times U(N_F)$ 
supplemented by the Chern-Simons terms and massive self-dual rank-two field $B_{mn}= B_{+,mn} + i B_{-,mn}$
and massive complex scalar $ X= X_{+} + i X_{-}$ both in the
bifundamental representation of the gauge group. The interaction terms in the Lagrangian involving the
rank-two term looks as follows
\beq
L_{int}= \lambda_{QCD} X_{\pm}F_{V,mn}B^{mn}_{\pm},
\eeq
where $F_V$ is the gauge curvature for the vector gauge field $A_V = A_L +A_R$.
The non-vanishing coefficient $\lambda_{QCD}$ in front of triple interaction term XBF implies the non-vanishing 
magnetic susceptibility $\chi_F$ of the quark condensate. It was shown in \cite{vainshtein,gk} 
that the non-vanishing susceptibility also follows from the anomaly in the axial current which
corresponds to the nontrivial Chern-Simons term in the holographic action. In the PCAC 
approximation $\chi_F= -\frac{N_c}{4 \pi^2 f_{\pi}^2}$ \cite{vainshtein} and one could say that the external field 
induces the spin polarization of the condensate via the Goldstone modes.

\section{Condensate polarization in  magnetic field}

To calculate the condensate of $<\psi\sigma_{\mu\nu}\psi>$ induced by the external magnetic field we will use the following procedure. First we consider the behavior of the  field $B_{12}$ in the pure $AdS_4$ space without presence of external electromagnetic field. After that we switch on electromagnetic field $F_{12} = B$ and calculate how the solution for $B_{12}$ has changed. The susceptibility can be read off from the asymptotic behavior of this new solution. To make calculations easier we suppose that $r_0 = 0 = T$ and change to the variable $ z =\frac{1}{r}$.  From the dimensional analysis we choose $m^2=-\frac{6}{L^2}$ and the equation of motion for $B_{12}$ is 
	\beqn
	\partial_z \left[z^2\partial_z B_{12}\right] + m^2  B_{12}= \lambda \Psi F_{12},\quad
	A_0 = \mu - \rho z.\label{1}
	\eeqn
	This equation has the following solution
	\beq
	B_{12} = C_1 z^2 + \frac{C_2}{z^3}.
	\eeq
	where $C_1  = <\psi\sigma_{12}\psi>$ and $C_2$ is a source for this operator in the
	boundary theory. 
	After that we switch on an external magnetic field on the right-hand side of the equation (\ref{1})
	\beqn
	\Psi = z^2 <\psi\psi> + \mathcal{O}(z^3),\quad z\to 0\notag\\ 
	F_{12} = B + \mathcal{O}(z^2),\quad z \to 0,\notag\\
	\Psi F_{12} =  z^2 B <\psi\psi> + \mathcal{O}(z^3), \quad z \to 0.
	\eeqn
	This term modifies the solution in the following way
	\beq
	B_{12} = \left(C_1 - \frac{\lambda}{5} B\log z\right) z^2,
	\eeq
	that we can attribute the new additional term to the
	non-vanishing magnetic susceptibility of the condensate. It gives us the following expression for susceptibility 
	\begin{equation}
	<\psi \sigma_{\mu\nu} \psi> = - \frac{\lambda}{5} \log \frac{z_{UV}}{z_0} <\psi\psi> F_{\mu\nu}, 
	\end{equation}
	where $z_{UV}$ is a UV cutoff and $z_0$ is some  IR scale. 
	
	The IR scale $z_0$ entering the logarithm deserves some explanation. There is some natural scale
	in the holographic approach which yields the scale of the scalar condensate. It is related
	to the parameter of the gravity solution $r_0$ however to identify it more precisely it is necessary
	to perform more refined analysis. On the other hand it is possible to get some intuition concerning
	this scale if we consider the inhomogeneous external magnetic field.

	Hence we look at  plane wave for magnetic field $F_{12} = B \exp\left(- i \omega t + i \vec{k} \vec{x}\right)$. The equation of motion for the field $B_{12}(z)$ reads as
	\beqn
	\partial_z \left[z^2\partial_z B_{12}\right] + (m^2 + p^2 z^2)  B_{12}= \lambda \Psi F_{12},\notag\\
	{\rm where}\quad p^2 = \omega^2 - k_1^2 - k_2^2.
	\eeqn
	and has the following solution in the limit $z \to 0$
	\beqn
	B_{12}(z) = -\frac{\lambda}{5}\, F_{12}\, <\psi\psi>\, \log (p a)\, z^2,\quad z \to 0.
	\eeqn
	This yields the following formula for the condensate 
	\beq
	<\psi \sigma_{\mu} \psi> = -\frac{\lambda}{5}\log(p a) B_\mu <\psi\psi>,
	\eeq
	where $a$ is a microscopic scale for superconductor (e.g. an interatomic distance)
	and $\sigma_{\mu}=\frac{1}{2}\epsilon_{\mu\nu\rho}\sigma_{\nu\rho},\quad 
	B_{\mu}=\frac{1}{2}\epsilon_{\mu\nu\rho}B_{\nu\rho}$.

	\section{Why is $\lambda$ not equal to zero?}
	
	In this Section we shall present the semi-qualitative arguments in favor of $\lambda\neq 0$. 
	First, comment on the derivation of the similar constant in QCD . It has been evaluated from the 
	correlator of the tensor and vector
	currents $<VT>$ in the boundary theory \cite{karch,harvey} which
	yields at large $Q^2$ in the massless QCD
	\beq
	\int dx e^{iQx}<V_{\mu}(0)T_{\nu \rho}(x)> \propto <\bar{\Psi} \Psi>(\eta_{\mu\nu} Q_{\rho} 
	-\eta_{\mu\rho} Q_{\nu}) Q^{-2}.  + O(Q^{-4}).
	\eeq
	where the Euclidean OPE of two currents is taken into account. 
	The correlator at low virtualities is saturated by the $\rho$-meson state which has 
	non-vanishing residues both for vector and tensor currents.

	On the other hand the same correlator can be evaluated holographically
	using the standard recipe that is varying the classical action 
	in the bulk theory over the boundary values of the 
	vector and tensor fields
	\beq
	<V(0)T(x)> = \frac{\delta^2 S_{cl}}{\delta A(0) \delta B(x)}.
	\eeq
	The bulk term $\lambda_{QCD} XBF$ contributes and 
	taking into account the boundary behavior of the neutral scalar $X(z) = <\bar{\Psi} \Psi>z^3 +\dots $ one gets
	upon comparison of two expressions for $\frac{<\bar{\Psi} \Psi>}{Q^{2}}$ terms
	\beq
	\lambda_{QCD}= -\frac{3N_c}{4\pi^2}.
	\eeq
	
	In our case similar calculation goes as follows. Consider the correlator $<VT_c>$
	once again and hunt for the $\frac{<\psi \psi>}{Q^{2}}$ term assuming that $Q^2$ is large 
	enough. The charged tensor current looks as $\psi \sigma \psi$ and
	two fermion legs can be send to the condensate yielding the
	non-vanishing contribution from the tree diagram
	\beq
	\int dx e^{iQx}<V_{\mu}(0)T_{\nu \rho}(x)> \propto <\psi \psi>(\eta_{\mu\nu} Q_{\rho} 
	-\eta_{\mu\rho} Q_{\nu}) Q^{-2} + O(Q^{-4})\label{correlator}
	\eeq

	At the bulk side we focus at the $\Psi BF$ term again and take into account 
	the boundary behavior of the charged  scalar $\Psi(z) =  z^2 <\psi \psi> + O(z^3) $.
	Evaluating the bulk action with this boundary condition we get the contribution proportional
	to the s-wave condensate as well. Equating the leading terms in correlators evaluated in the bulk and 
	in the boundary superconductor  we obtain that $\lambda \neq 0$. However at the holographic side 
	we have the additional factor $\log Q^2$ which obstructs the estimate of the numerical 
	value of $\lambda$.
	
Here we present also  similar calculation for $3+1$ superconductor when the dual action reads as  
	\begin{gather}
	S = \int d^5 x \sqrt{-g}\left[R + \frac{\sigma}{L^2}-\frac14 F_{\mu\nu}^2 + \left|D_\mu X\right|^2 - m_\Psi^2 |X|^2 + \left|d B - i A \wedge B\right|^2 - m_B^2 |B|^2\right.\notag\\\left. + \lambda X F_{\mu\nu} B^{\dagger,\mu\nu}+\lambda X^\dagger F_{\mu\nu} B^{\mu\nu}\right],\notag\\
	ds^2 = \frac{1}{z^2}\left[-dt^2+dx_i^2 + dz^2\right],\quad A_0 = \mu - \rho z^2.
	\end{gather}
	where $m_B$ and $m_\Psi$ are chosen to satisfy tree-level dimensions for dual operators on the boundary
	\begin{gather}
	z^5 \partial_z\left( z^{-3} \partial_z\right) X + m_\Psi^2 X = 0,\quad X \sim  z^3<\psi\psi> + z J_\psi,\quad m_\Psi^2 = 3, \notag\\
	z \partial_z \left(z \partial_z\right) B_{12} + m_B^2 B_{12} = 0,\quad B_{12} \sim z^3<\psi \sigma_{12}\psi> + \frac{J_B}{z^3}, \quad m_B^2 = -9,
	\end{gather}
	and we have set the radius of AdS to be $L=1$. We can calculate the correlator of tensor and electric current $<VT>$ imposing the following boundary conditions for the fields in AdS
	\begin{gather}
	B_{12} = \frac{J_B(x)}{z^3}, \quad
	A_1 = J_1(x) z^2, z \to 0.
	\end{gather}
	If we assume that $J_1,J_B \sim e^{\pm i p x}$ we get the following equations for $B_{12}$ and $A_1$ fields
	\begin{gather}
	z \partial_z\left( z^{-1} \partial_z\right) A_1 + p^2 A_1 = 0, \notag\\
	z \partial_z \left(z \partial_z\right)B_{12} +\left(m_B^2+p^2 z^2\right) B_{12} = 0.
	\end{gather}
	The solutions can be written as  linear combinations of Bessel functions with proper boundary conditions
	\begin{gather}
	B_{12} = - \frac{\pi p^3}{16}J_B\, e^{i p x} \, Y_3(p z), \quad A_1 = \frac{2 z}{p} J_B J_1(p z) e^{-i p x}. 
	\end{gather}
	That yields the following expression for the correlator
	\begin{gather}
	<V(-p) T(p)> = - \frac{\lambda \pi p_2}{8} p^2 \int\limits^\infty_0  dz J_1(pz) Y_3(pz) X(z) = -\frac{\lambda \pi p_2}{8} p^2 \int^\infty_0 dw J_1(w) Y_3(w) X(w/p) \approx_{p\gg 1} \notag\\\approx_{p\gg 1} - \frac{\lambda \pi p_2}{8 p^2} <\psi\psi> \int \limits^\infty_0 dw w^3 J_1(w) Y_3(w).
	\end{gather}
	Comparing this answer with $\eqref{correlator}$ we get that $\lambda \neq 0$ for $3+1$ case as well.

	\section{On the vortical susceptibility of the Cooper condensate}

Let us make a few comments concerning the similar response of 
the s- wave supercunductor on the rotation. We introduce
the corresponding susceptibility  postponing the holographic
study for the separate publication.
	
There is a well-known analogy between an external magnetic field and
the rotating frame, manifested 
in the non-relativistic case by the substitution $e_f \vec{B} \leftrightarrow m \vec{\Omega}$ ( see, for example \cite{svistunov}).  It is thus natural 
to introduce a response of the Cooper  condensate in superconductor to the rotation
which can be parametrized as follows 
\beq
\label{vort_s}
<0|\Psi\sigma_{\mu\nu}\Psi|0>= \chi_{s,G} <\Psi\Psi>G_{\mu\nu}.
\eeq
We treat the rotation via the curvature of an external graviphoton field  $G_{\mu\nu}$
and denote the corresponding vortical susceptibility 
of the s-wave Cooper condensate as $\chi_{s,G}$.

Let us recall that the graviphoton field
is introduced as the specific form of the background metric
\beq
ds^2=(1+ 2\phi_g)dt^2 -(1-2\phi_g)d^2\vec{x} + 2\vec{A_g}d\vec{x}dt.
\eeq
The gravimagnetic field corresponds to the angular velocity of rotation at small velocity  
\beq
\vec{B}_g\propto \vec{\Omega},
\eeq
however at large velocities the relation between the gravimagnetic field and 
the angular velocity is more complicated.

In the rotating superconductor the magnetic field in the bulk is generated
\beq
\vec{B}= -\frac{2m \vec{\Omega}}{e},
\eeq
As we have shown in the previous Section  this magnetic field  induces the triplet component in the material
with non-vanishing susceptibility. Hence this argument suggests that p-wave component is also
genetated  in the s-wave condensate under rotation. This can be considered in the 2+1 case 
where such component can be generated in the plane everywhere besides the droplets where
the condensate vanishes.

We postpone the holographic analysis of the rotating case when the angular velocity is
introduced via the rotating black hole. The analysis  is expected to be parallel to the
discussion in \cite{afgk}.

\section{Conclusion}

In this Letter we discuss in the holographic framework the effect of the external magnetic field  on the s-wave superconductor at small temperature. We argue that apart from the conventional Meissner effect
there is the possibility of  generation of the homogeneous p-wave component  in the volume of superconductor due to the polarization of the Cooper condensate. The magnetic and rotational susceptibilities of the Cooper condensate are 
introduced. We consider the linear approximation when the triplet component is proportional to the external field and argue
that the susceptibilities do not vanish.
	
In the usual setting the external magnetic field influences the  volume of superconductor via the Abrikosov vortices and the condensate vanishes at their cores. The effect of condensate polarization we consider seems to have nothing to do with the vortices since we do not assume the vanishing of the s-wave condensate anywhere. The appearance of the p-wave admixture in s-wave superconductor caused by the Rashba term or by antiferromagnetic component seems to be	the most related phenomena. It would be also important to fit our 
observation with discussion in \cite{p+s,kiritsis,cai}.
	
Certainly it is interesting to make the numerical estimates of the magnetic and vortical susceptibilities of s-wave SC and take into account the temperature dependence. It would be also interesting to investigate
the susceptibilities of the exciton condensate in the external fields which is more close analogue to QCD case.
	
We are grateful to David Huse, Alexander Krikun, Phuang Ong  and Nicolai Prokof'ev for the useful comments.
The work by Fedor Popov has been supported by the RScF grant 16-12-10151. The work of A.G. is
supported in part by the grant RFBR- 15-02-02092   and Basis Foundation fellowship.

\end{document}